\documentclass[twocolumn,aps,prb,floats,superscriptaddress,showpacs]{revtex4}
\usepackage{amsmath}
\usepackage{epsfig}
\usepackage{graphicx}
\usepackage{dcolumn}
\usepackage{bm}
\usepackage{color}

\def\gapp{\lower.35em\hbox{$\stackrel{\textstyle>}{\sim}$}}
\def\lapp{\lower.35em\hbox{$\stackrel{\textstyle<}{\sim}$}}

\begin{document}

\title{{Tunable Casimir equilibria with phase change materials: from quantum
trapping to its release}}
\author{Lixin Ge}
\email{lixinge@hotmail.com}
\affiliation{School of Physics and Electronic Engineering, Xinyang Normal University,
Xinyang 464000, China}
\author{Xi Shi}
\affiliation{Department of physics, Shanghai Normal University, Shanghai, 200234, China}
\author{Zijun Xu}
\affiliation{School of Physics and Electronic Engineering, Xinyang Normal University,
Xinyang 464000, China}
\author{Ke Gong}
\affiliation{School of Physics and Electronic Engineering, Xinyang Normal University,
Xinyang 464000, China}
\date{\today}

\begin{abstract}
A stable suspension of nanoscale particles due to the Casimir force is of
great interest for many applications such as sensing, non-contract
nano-machines. However, the suspension properties are difficult to change
once the devices are fabricated. Vanadium dioxide (VO$_2$) is a phase change
material, which undergoes a transition from a low-temperature insulating
phase to a high-temperature metallic phase around a temperature of 340 K. In
this work, we study Casimir forces between a nanoplate (gold or Teflon) and
a layered structure containing a VO$_2$ film. It is found that stable
Casimir suspensions of nanoplates can be realized in a liquid environment,
and the equilibrium distances are determined, not only by the layer
thicknesses but also by the matter phases of VO$_2$. Under proper designs, a
switch from quantum trapping of the gold nanoplate (``on" state) to its
release (``off" state) as a result of the metal-to-insulator transition of VO%
$_2$, is revealed. On the other hand, the quantum trapping and release of a
Teflon nanoplate is found under the insulator-to-metal transition of VO$_2 $%
. Our findings offer the possibility of designing switchable devices for
applications in micro-and nano-electromechanical systems.
\end{abstract}

\maketitle

%\pacs{78.67.Pt, 73.20.Mf, 42.25.-p, 41.20.Jb}

\section{Introduction}

Micro- and nano-electromechanical systems (MEMS and NEMS), which integrate
electrical and mechanical functionality on the micro- and nano-scales, have
attracted enormous attention \cite{Lys:18,Cra:00}. Thanks to small sizes,
the MEMS and NEMS exhibit low mass, high mechanical resonance frequencies
and quantum effects, leading to a broad range of applications such as
biological/chemical detections \cite{Eom:11}, accelerometers \cite{Xu:11}
and micro/nanomachines \cite{Wan:13}. One major problem in MEMS and NEMS is
the $stiction$ which makes the systems collapse and permanent adhesion
caused by the attractive Casimir forces \cite{Buk:01, Cha:01}. The Casimir
force is a macroscopic quantum effect which arises from quantum fluctuations
of the electromagnetic field \cite{Cas:48}. In most cases, two neutral,
parallel plates consisted of the same materials are attractive to each
other, and the magnitudes of the attraction depend on several parameters such as separations, geometric thicknesses, finite conductivities and temperatures (see, e.g., the review \cite{Kli:09} and Refs.\cite{Yam:08, Yam:10}. Therefore, repulsive Casimir forces are highly required for
non-contact and low-friction MEMS and NEMS. The repulsive Casimir forces
have been intensively studied in many systems \cite{Woo:16} including
liquid-separated environments \cite{Mun:09,Van:10,Pha:11,Dou:14},
meta-materials \cite{Ros:08,Zha:09,Zha:11,Son:18}, topological insulators
\cite{Gru:11, Che:12, Nie:13} and specific geometrics \cite{Tan:17,Lev:10}.
In addition, the concept of Casimir equilibria was also investigated, using
the enclosed geometries \cite{Rod:08,Rah:10} and dispersive materials \cite%
{Rod:10}. Lately, stable Casimir equilibria of nanoplates above a
Teflon-coated gold substrate were reported by Zhao et al \cite{Zha:19}.
However, the Casimir equilibria of previous studies were mainly in passive
systems. Once the devices are fabricated, the trapping properties are
difficult to change. Thus, the tunable trapping or even the switching from
the trapping to its release by external stimuli (e.g., heating, electric
fields or optical waves) is highly desired in MEMS and NEMS.

In order to active modulate the Casimir effect, one straight way is to
change the dielectric properties of materials under external means \cite%
{Tor:12,Sed:13, Tor:10}. Vanadium dioxide (VO$_2$) \cite{Sha:18, Zyl:75} is a
phase change material(PCM), which undergoes a transition from a
low-temperature insulating phase to a high-temperature metallic phase at
critical temperature 340 K. The phase transition of VO$_2$ is accompanied by
a structural transformation from the monoclinic phase to the tetragonal one.
Meanwhile, the dielectric function of VO$_2$ changes dramatically during the
phase transition, leading to many interesting applications \cite{Wu:17,
Liu:17, Kat:12, Van:12}. In general, the phase transition of VO$_2$ can be
induced by changing the temperature of systems. Alternatively, the phase
transition can be driven by optical lasers \cite{Cav:01, Rin:08} or
electrical gratings \cite{Qaz:08, Nak:12} on a sub-picosecond timescale.
Recently, VO$_2$ has been employed to study the tunable Casimir effect in
the vacuum \cite{Gal:09, Pir:08,Cas:07}.  For a large separation (e.g., $>$1 $\mu$m), the contrast of Casimir forces due to the phase-transition is quite large (e.g., over 2 times for two semi-infinite plates of VO$_2$, this value could be even larger for the case of finite thickness \cite{Gal:09, Pir:08}). As the separation is small (e.g., $\sim$100 nm), however, the modulation of Casimir forces owning to the phase transition and finite-thickness decreases greatly \cite{Pir:08, Cas:07}. Nonetheless, the Casimir forces are always attractive and only magnitude modulations have been reported in a vacuum-separated configuration. The influences of phase transition of VO$_2$ on the sign modulation of Casimir forces (e.g., from attraction to repulsion) are yet less explored. In a liquid environment, the function of sign modulation and the related phenomena such as tunable Casimir equilibria are expected based on the phase transition of VO$_2$.

Here, the Casimir forces between a nanoplate and a layered structure
separated by a liquid are investigated. The layered structure consists of
two kinds of materials, i.e., Vanadium dioxide (VO$_2$) and Teflon. It is
found that stable Casimir equilibria of gold nanoplates can be realized when
a VO$_2$ film is buried under a semi-infinite Teflon. The properties of
Casimir equilibria are determined, not only by the layer thicknesses but
also by the matter phases of VO$_2$. For thick-film VO$_2$, the Casimir
equilibria and quantum traps can be achieved for both the metallic and
insulating phases. On the other hand, a switch from quantum trapping of the
gold nanoplate(``on'' state) to its release (``off'' state) can be triggered
by the metal-to-insulator phase transition when the thickness of VO$_2$ is
thin (e.g., 20 nm). Finally, stable suspensions of Teflon nanoplates are
also proposed with a complementary design, where the Teflon substrate is
coated by a VO$_2$ film. Unlike the case of gold nanoplates, the quantum
trapping of Teflon nanoplates and its release correspond to the insulating
and metallic phases of VO$_2$. Moreover, the switching phenomena can be
realized only with a several-nanometers thickness of VO$_2$.

\begin{figure}[tbp]
\centerline{\includegraphics[width=8.4cm]{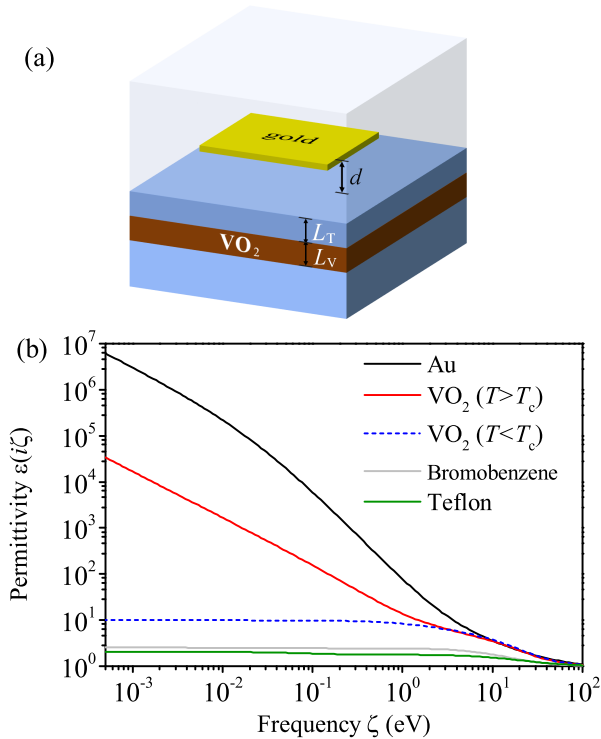}}
\caption{(color online) (a) Schematic view of a gold nanoplate suspended in
a liquid environment. (b) The permittivity of different materials (gold, VO$%
_2$, bromobenzene and Teflon) as a function of imaginary frequency. }
\label{Fig:fig1}
\end{figure}

\section{Theoretical models}

The system in this work is schematically shown in Fig. 1(a), where a gold
nanoplate with thickness $L_{g}$ is suspended in a liquid of bromobenzene.
The separation between the nanoplate and the substrate is $d$. The substrate
is composed of a VO$_{2}$ film buried under a semi-infinite plate of Teflon.
The thicknesses of the top-layer Teflon and VO$_{2}$ are denoted as $L_{T}$
and $L_{\mathrm{V}}$, respectively. The in-plane dimension of the gold
nanoplate is much larger than $L_{g}$ and $d$, and it is considered as a
slab during our calculations. The Casimir force is calculated by $%
F_{c}=-\partial E_{c}(d)/\partial d$, where $E_{c}(d)$ is the Casimir energy
between the gold nanoplate and the substrate, having the form \cite{Nie:13,
Zha:19}
\begin{equation}
E_{c}(d)=A\hbar \int_{0}^{\infty }\hspace{-1pt}\frac{d\xi }{2\pi }\int \frac{%
d^{2}\mathbf{k_{\Vert }}}{(2\pi )^{2}}\log \det \left[ 1-\mathbf{R_{1}}\cdot
\mathbf{R_{2}}e^{-2k_{3}d}\right] ,
\end{equation}%
where $\hbar $ is the reduced Planck constant, $A$ is the in-plane area, $%
\mathbf{k_{\parallel }}$ is the parallel wavevector, $k_{3}=\sqrt{%
k_{\parallel }^{2}+\varepsilon _{liq}(i\xi )\xi ^{2}/c^{2}}$ is the vertical
wavevector, $c$ is the speed of light in vacuum, $\varepsilon _{liq}(i\xi )$
is the permittivity of the intervening liquid evaluated with imaginary
frequency $\omega =i\xi $, $\mathbf{R}_{1,2}$ is the $2\times 2$ reflection
matrix for layered structures, having the form
\begin{equation}
\mathbf{R_{j}}=\left(
\begin{array}{cc}
r_{j}^{s} & 0 \\
0 & r_{j}^{p}%
\end{array}%
\right) ,
\end{equation}%
where $r_{j}$ with $j$=1 and $j$=2 are the reflection coefficients for the
upper and lower layered structures, and the superscripts $s$ and $p$
correspond to the polarizations of transverse electric ($\mathbf{TE}$) and
transverse magnetic ($\mathbf{TM}$) modes, respectively. Note that the temperature $T$ for Eq. (1) equals 0 K and it is an effective
approximation as the separation $d$ is smaller than 1 $\mu m$ for finite
temperatures \cite{Mil:04}. For a nanoplate
suspended in a liquid, the reflection coefficients can be given analytically
as follows \cite{Zha:11}
\begin{equation}
r^{\alpha }=\frac{r_{0,j}^{\alpha }+r_{j,0}^{\alpha }e^{-2K_{j}L_{j}}}{%
1+r_{0,j}^{\alpha }r_{j,0}^{\alpha }e^{-2K_{j}L_{j}}},
\end{equation}%
where $\alpha =s$ and $p$, $L_{j}$ is the thickness of the nanoplate, $K_{j}=%
\sqrt{k_{\parallel }^{2}+\varepsilon _{j}(i\xi )\xi ^{2}/c^{2}}$ is the
vertical wavevector, $\varepsilon _{j}(i\xi )$ is the permittivity of the
nanoplate.  The subscripts of $r_{m,n}^{\alpha }$ represent the light is
incident from the medium $m$ to $n$ (0 means the liquid).

%and the coefficients $r_{mn}^{\alpha}$ are written as
%\begin{equation}
%\begin{array}{c}
%r_{mn}^{s}=(K_{m}-K_{n})/(K_{m}+K_{n}) \\
%r_{mn}^{p}=(\varepsilon _{n}K_{m}-\varepsilon _{m}K_{n})/(\varepsilon
%{n}K_{m}+\varepsilon _{m}K_{n})%
%\end{array}%
%\end{equation}

Alternatively, the reflection coefficients for layered structures can be
calculated by a transfer matrix method. The general form is given as $%
r=M_{21}/M_{11}$, where $M_{21}$ and $M_{11}$ are the elements of the $M$
matrix\cite{Zha:13}. The $M$ matrix is the multiplications of transmission
matrices across different interfaces and propagation matrices in different
layers. Considering an arbitrary $N$-layer system, the $M$-matrix is given
as :
\begin{equation}
M=D_{0,1}P(L_{1})D_{1,2}P(L_{2})...D_{N-1,N}P(L_{N})D_{N,N+1},
\end{equation}%
where the transmission matrix $D_{j,j+1}$ is given as:
\begin{equation}
D_{j,j+1}=\frac{1}{2}\left[
\begin{array}{cc}
1+\eta & 1-\eta \\
1-\eta & 1+\eta%
\end{array}%
\right],
\end{equation}%
where $\eta =\varepsilon _{j}(i\xi)K_{j+1}/(\varepsilon _{j+1}(i\xi)K_{j})$
for p-polarization and $\eta =K_{j+1}/K_{j}$ for s-polarization. The
propagation matric in the $j$-th layer (for both $s$ and $p$ polarizations)
is written as:
\begin{equation}
P(L_{j})=\left[
\begin{array}{cc}
e^{K_{j}L_{j}} & 0 \\
0 & e^{-K_{j}L_{j}}%
\end{array}%
\right].
\end{equation}
For example, we have $N=2$ for the multilayered substrate in Fig. 1. The $M$
matrix is given by $M=D_{0,1}P(L_{1})D_{1,2}P(L_{2})D_{2,3}$, where the
subscripts 0, 1, 2 and 3 represent the media of liquid, Teflon, VO$_2$ and
Teflon (from top to down); the thicknesses $L_{1}=L_T$, $L_2=L_V$.

\section{Results and discussions}

Figure 1(b) shows the permittivity for different materials, where the used models and parameters are given in the Appendixes. The dielectric
function of VO$_{2}$ changes dramatically under different temperatures. For
temperature $T>T_{c}$, VO$_{2} $ is in the metallic phase and it acts as a
poor metal. For $T<T_{c}$, it is in the insulating phase (or called
semiconducting phase), and the corresponding dielectric function nearly
matches that of intrinsic silicon at low frequency \cite{Pir:08}. To create
repulsive Casimir forces between two dissimilar plates separated by a
liquid, the permittivity should satisfy $\varepsilon _{1}(i\xi )>\varepsilon
_{liq}(i\xi )>\varepsilon _{2}(i\xi )$ for a vast range of frequency \cite%
{Mun:09}. Clearly, the dielectric functions of gold and VO$_{2}$ (either
metallic or insulating phase) are larger than that of bromobenzene over a
wide range of frequency. Therefore, the Casimir force is always attractive
for the layered structure of gold/bromobenzene/VO$_{2}$. While the Casimir
force for the structure of gold/bromobenzene/Teflon is repulsive instead.
Nonetheless, the Casimir equilibria can not be found for above two layered
structures.

\begin{figure}[tbp]
\centerline{\includegraphics[width=8.5cm]{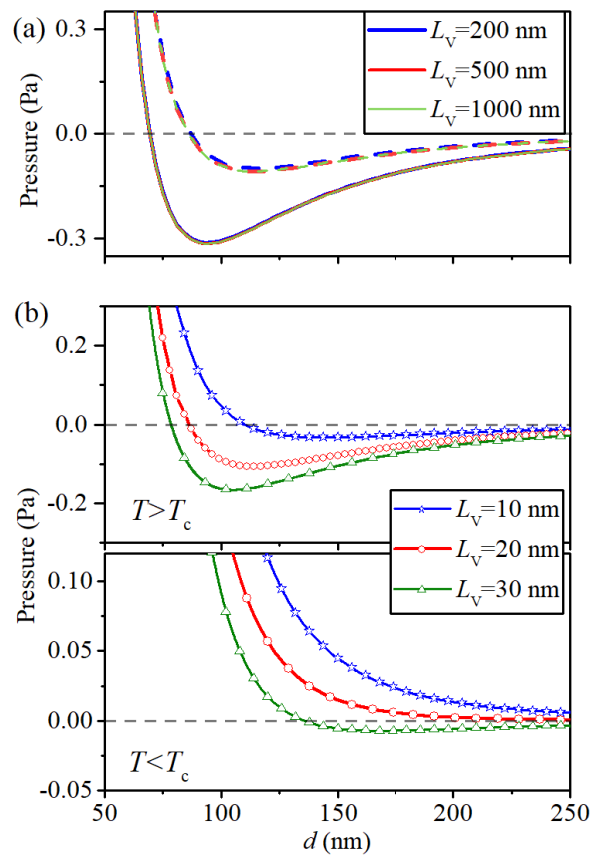}}
\caption{(color online) Casimir pressure via different thicknesses of VO$_2$%
, where the thickness $L_T$=45 nm and $L_g$=40 nm are fixed. (a) Thick
films. The solid and dashed lines represent the pressure for the metallic
and insulating phases of VO$_2$, respectively. (b) Thin films. The positive
(negative) sign of the pressure corresponds to the repulsive (attractive)
force.}
\label{Fig:fig2}
\end{figure}

\subsection{Tunable Casimir equilibria for gold nanoplates}

Now we consider the Casimir forces as the substrate is composed of a VO$_2$
film and Teflon (see Fig. 1(a)). The Casimir pressure ($P_c=F_c/A$) for the
thick film of VO$_2$ is given in Fig. 2(a). The results show that the curves
are almost identical for $L_\mathrm{V}$=200, 500 and 1000 nm, indicating the
weak impact of the thickness for thick-film configurations. The pressure is
repulsive at small separation (e.g., $d<60$ nm), making the nanoplate stay
away from the substrate. As the separation increases further, the Casimir
equilibria (zero pressure) occur and quantum traps can be realized for both
metallic (solid lines) and insulating phases (dashed lines). In
addition, the equilibrium distance $d_c$ is shifted under the phase
transition of VO$_2$. On the other hand, the thin-film thickness and the
phase transition of VO$_2$ can play an important role in Casimir pressure as
shown in Fig. 2(b). For the thickness $L_\mathrm{V}$ =10 and 20 nm, quantum
traps can be realized for the metallic phase, whereas no trap is found for
the insulating phase. Under such configurations, a switch from quantum
trapping of the nanoplate(``on" state) to its release (``off" state) can be
triggered by the metal-insulator transition of VO$_2$. However, the quantum
trapping occurs for both metallic and insulating phases as the thickness $L_%
\mathrm{V}$ increases to 30 nm, and the ``off" state disappears. Compared with the vacuum-separated configuration \cite{Cas:07}, not only the magnitude of Casimir forces can be modified in a liquid environment, but also the sign could be switched (e.g., from attraction to repulsion for $d$=100 nm, $L_V$=30 nm), due to the phase-transition of VO$_2$.

\begin{figure}[tbp]
\centerline{\includegraphics[width=8.4 cm]{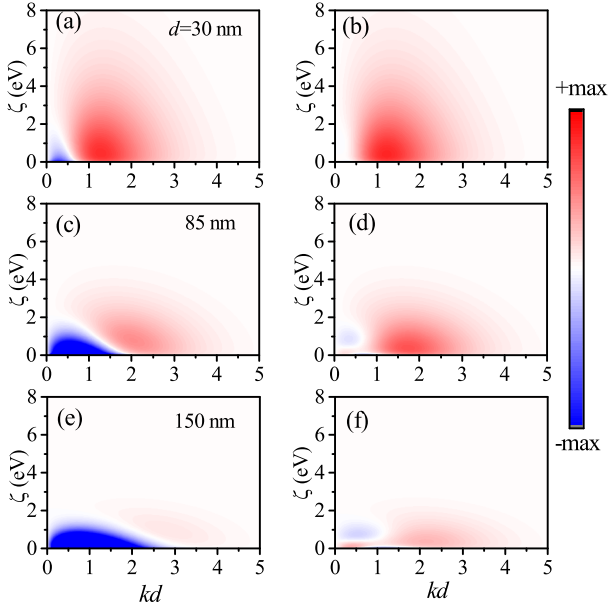}}
\caption{(color online) Casimir pressure contributed from different
frequencies and different parallel wavevectors. (a) and (b) $d$=30 nm; (c)
and (d) $d$=85 nm (close to critical separation); (e) and (f) $d$=150 nm.
(a), (c)and (e) VO$_2$ in the metallic phase ($T>T_c$); (b), (d) and (f) VO$%
_2$ in the insulating phase ($T<T_c$). The layer thicknesses are set as $L_%
\mathrm{V}$=20 nm and $L_T$=45 nm. }
\label{Fig:fig3}
\end{figure}

To understand the switch transition from the ``on" to the ``off" state, the
contour plots of Casimir pressure are shown in Fig. 3 under different
separations. The sign of pressure is determined by the competition of VO$_2$
film (attraction) and low-refractive-index Teflon (repulsion). For small
separation $d$=30 nm, the pressure is dominant by the repulsive component as
shown in Figs. 3(a) and 3(b). For the metallic phase, the attractive
component increases and it compensates the repulsive one as the separation
becomes 85 nm ($d\approx d_c$), resulting in Casimir equilibrium (see Fig.
3(c)). While the repulsion is still dominant for the insulating phase as
shown in Fig. 3(d). As $d$ increases further to 150 nm, the Casimir pressure
turns out to be dominantly attractive in Fig. 3(e) for the metallic phase,
resulting in a restoring force for stable trapping. By contrast, the
pressure is still dominant by repulsion for the insulating phase as shown in
Fig. 3(f). The pressure maps between the metallic and insulating phases are
almost identical for large energy (e.g., $>$2 eV), whereas the discrepancy
manifests at low energy. The results indicate that the attractive component
appears only at low frequency and small $k$ vector for metallic VO$_2$,
where the field cannot penetrate the metal\cite{Zha:19}. Conversely, the
field can penetrate the thin-film of insulating VO$_2$ easily, leading to
repulsive Casimir forces.

\begin{figure}[tbp]
\centerline{\includegraphics[width=8.5cm]{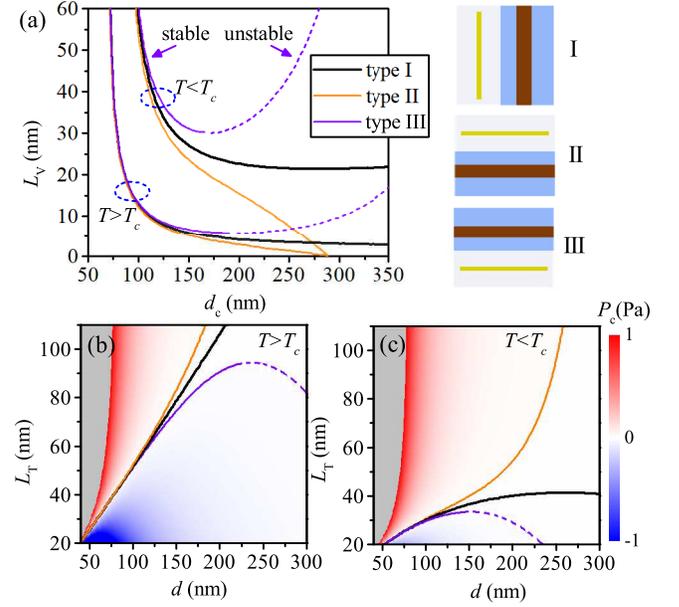}}
\caption{(color online) (a)The equilibrium distances via the thicknesses of
VO$_2$ under three different configurations (see the inset on the right).
The thickness $L_T$ is set as 45 nm. The solid (dashed) curves for type III
represent stable (unstable) equilibria. Contour plots of Casimir pressure
via the thicknesses of coating Teflon for (b) metallic VO$_2$ and (c)
insulating VO$_2$, where the thickness $L_\mathrm{V}$=20 nm is fixed. In (b)
and (c), the gray zones represent a strong repulsive pressure larger than 1
Pa. The colors of the curves denote the same meaning as those in (a).}
\label{Fig:fig4}
\end{figure}

Practically, the influences of gravitation and buoyancy on the force
balances should be taken into account. The condition for the force
equilibrium is written as $\vec{n}\cdot (\mathbf{F}_c+\mathbf{F}_{\mathrm{GB}%
})$=0, where $\vec{n}$ is the unit vector normal to the surface, $F_{\mathrm{%
GB}}=(\rho_g-\rho_{liq})gL_gA$ is the sum of gravity and buoyancy, $g$ is
the gravitational acceleration, $\rho_g\approx$19.3 g/cm$^3$ and $%
\rho_{liq}\approx$1.50 g/cm$^3$ is the density of gold and liquid
bromobenzene, respectively. The magnitude of $F_{\mathrm{GB}}/A$ is about
7.0 mPa as the thickness $L_g$=40 nm. Three types of configurations are
depicted in the inset of Fig. 4(a) for the cross-section views. The type I
configuration corresponds to a zero-projection (or weightlessness in
aerospace), where the switching from quantum trapping (metallic state) to
its release (insulating state) can be obtained as $L_\mathrm{V}$ in a proper
range, from about 2 to 22 nm. For type II configuration, the attractive $F_{%
\mathrm{GB}}$ can compensate the long-range repulsive Casimir force at large
$d$, leading to stable suspensions for both $T>T_c$ and $T<T_c$. However,
the equilibrium distances are different, and it can be inferred that the
stiffness of trapping for metallic phase is stronger than that of the
insulating phase. For type III configuration (a flipped down system), the
switching between trapping and its release can also be realized.
Interestingly, there are two equilibrium distances for this configuration.
It is not difficult to know that the smaller equilibrium distance (solid
lines) is stable, whereas the other one (dashed lines) with larger distance
is unstable to small perturbations in position. For both type II and III
configurations, the deviations from Type I become strong as $d_c$ is large.

In addition to the thickness of VO$_2$ film, the top-layer Teflon can also
play a significant role in the Casimir effect. The plots of Casimir pressure
via the thicknesses of the coating Teflon $L_T$ are shown in Figs. 4(b) and
4(c), where $L_\mathrm{V}$=20 nm is fixed. The results show that the
switching between quantum trapping and it release occurs only when $L_T$ is
larger than about 42 nm (no gravity). The larger the $L_T$, the larger of
the position for the Casimir equilibrium. As $L_T$ is smaller than 42 nm, the
equilibrium distance is also small, and quantum trappings can be realized
for both metallic and insulating phases. For comparison, the gravitation and
buoyancy are taken into account. Again, strong discrepancies among three
configurations occur as the equilibrium positions larger than about 150 nm,
resulting from the comparable magnitude of $F_{GB}$ and the Casimir force.
The impact of $F_{GB}$ can be further reduced by decreasing the thickness $%
L_g$ near the skin depth (about 22 nm) \cite{Lis:05}.

\begin{figure}[tbp]
\centerline{\includegraphics[width=8.5cm]{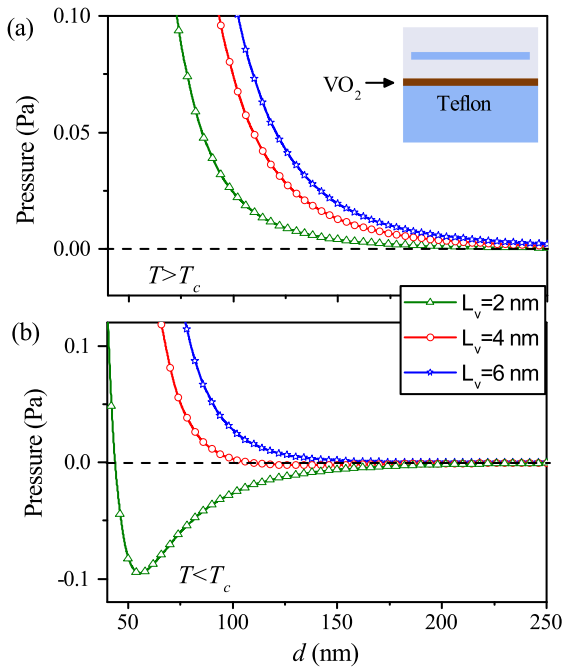}}
\caption{(color online) Casimir pressure for a complementary design. A thin
film of VO$_2$ with thickness $L_V$ is deposited on a Teflon substrate.
(a)The metallic VO$_2$. (b)The insulating VO$_2$. The thickness of the
suspended nanoplate is set as 100 nm. }
\label{Fig:fig5}
\end{figure}

\begin{figure}[tbp]
\centerline{\includegraphics[width=8.5cm]{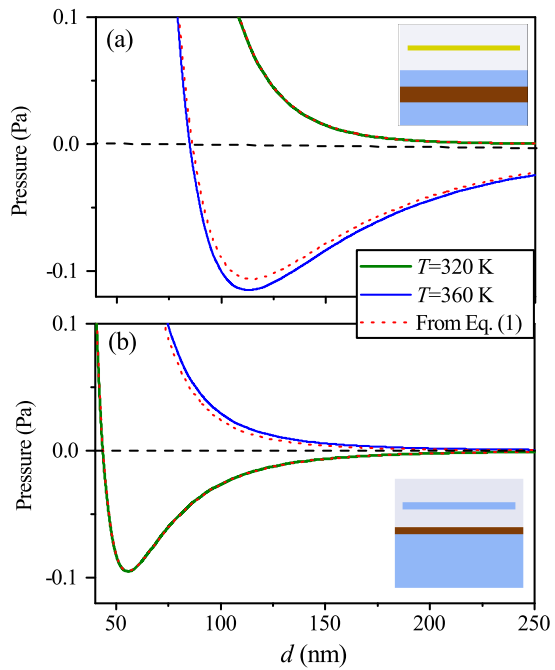}}
\caption{(color online) Casimir pressure calculated for finite temperatures
and 0 K approximation from Eq. (1). (a)The trapping and release of a gold
nanoplate. The parameters for the substrate are $L_T$=45 nm and $L_V$=20 nm.
(b)The trapping and release of a Teflon nanoplate. The thickness $L_V$ is
set as 2 nm. }
\label{Fig:fig6}
\end{figure}

\subsection{Tunable Casimir equilibria for Teflon nanoplates}

The active control of the low-refractive-index nanoplates can also be
significant in many applications. Inspiring by the work \cite{Zha:19}, a
complementary design is schematically shown in the inset of Fig. 5(a). A
Teflon nanoplate is suspended in a liquid of bromobenzene, and the substrate
is a semi-infinite plate of Teflon coated by a VO$_2$ film (high refractive
index). Under such design, the Casimir force is repulsive at very short
separation, due to the dominant interaction between Teflon/bromobenzene/VO$%
_2 $. As the separation increases, the attractive interaction from
Teflon/bromobenzene/Teflon can be dominant instead, resulting in a stable
Casimir trapping. To verify the design, the Casimir pressure is given
quantitatively in Figs. 5(a) and 5(b) as a function of separation.
Interestingly, the Casimir pressure shows a long-range repulsive behavior
for the metallic VO$_2$, which corresponds to the ``off" state. The
repulsion pressure becomes stronger as the thickness $L_\mathrm{V}$ enlarges
from 2 to 6 nm. For $L_\mathrm{V}$= 2 nm, a Casimir equilibria and strong
restoring forces can be found when VO$_2$ is in the insulating phase.
Therefore, the quantum trapping and release of a Teflon nanoplate can be
achieved under the insulator-to-metal transition of VO$_2$. As the thickness
is 4 nm, the restoring force decreases and the trapping stiffness drops
considerably. The calculation results indicate that the Casimir pressure is
quite sensitive to the thickness of VO$_2$. Due to the low density of Teflon
(2.1 g/cm$^3$), the pressure $F_{GB}/A$ for the Teflon nanoplate is about 0.6
mPa, which is reduced significantly compared with those of gold nanoplates.

\subsection{Finite temperatures effect}

To achieve the phase transition of VO$_2$, the temperatures of the devices
need to be changed. We assume that the dielectric functions of the gold and
Teflon are temperature-independent. For organic liquids, the change of
refractive index due to the temperature \cite{Li:94} is an order of $%
10^{-4}/ $ K, and the permittivity of bromobenzene is also treated as
temperature-independent. Nonetheless, it is interesting to check the finite
temperature effect on Casimir forces. The integral over frequency $\xi $ in
Eq. (1) now is replaced by a discrete summation \cite{Rah:09}:
\begin{equation}
\frac{\hbar }{2\pi }\int_{0}^{\infty }d\xi \leftrightarrow k_{b}T \overset{%
\infty }{\underset{n=0}{\sum}}^{\prime},
\end{equation}
where $\xi $ is replaced by discrete Matsubara frequencies $\xi _{n}=2\pi
\frac{k_{b}T}{\hbar }n(n=0,1,2,3\ldots ),$ $k_{B}$ is the Boltzmann's
constant and the prime denotes a prefactor 1/2 for the term $n$=0. The
Casimir pressures under different temperatures are shown in Figs. 6(a) and
6(b), where two different designs are demonstrated. It is found that the
curves for temperature 320 K (insulating phase) overlap with those
calculated from Eq. (1). For the temperature of 360 K, there is only a small
deviation between 0 K and 360 K. Overall, the calculation results from 320
and 360 K confirm the accuracy of the 0 K approximation. Recently, the
switching between repulsive and attractive Casimir forces based on PCM has
also been reported \cite{Bos:18}, where the equilibrium distances for
switching occur only at several nanometers. The equilibrium distances in our
work are more accessible to experiments, and it can be tuned by designing
the geometric thickness of VO$_2$ and the Teflon.

\begin{figure}[tbp]
\centerline{\includegraphics[width=8.2cm]{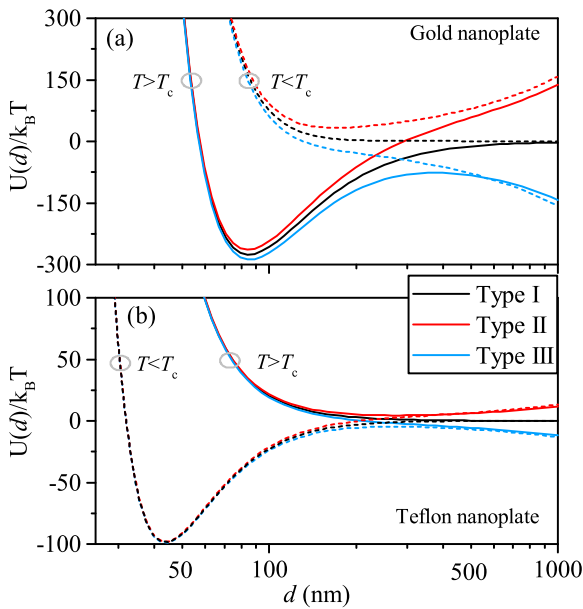}}
\caption{(color online) The total energy of a suspended gold nanoplate (a) and a Teflon nanoplate (b) under different types of gravity projection. The solid and dashed lines represent the cases for the metallic VO$_2$ ($T$=360 K) and insulating VO$_2$ ($T$=320 K), respectively. The in-plane area $A$ is set as 10 $\mu m \times$10 $\mu m$. Other parameters are kept the same as those in Fig. 6. }
\label{Fig:fig7}
\end{figure}
\subsection{The effect of Brownian motion}

In a real configuration, the position of a nanoplate has a fluctuation around the equilibrium distances due to the Brownian motion.  To evaluate the effect of Brownian motion, the total energy of the suspended nanoplate should be known, which are written as  $U(d)=E_c+\Lambda\times(E_g+E_b)$, where $E_c$ is the Casimir energy given by Eq. [1], $E_g=\rho_pgL_pAd$ and $E_b=-\rho_{liq} gL_pAd$ are respectively the energies caused by the gravity and buoyancy \cite{Pha:12}, $\rho_p$ and $L_p$ represent the density and thickness of the suspended nanoplate. The coefficient $\Lambda$ is the parameter depending on the gravity projection. For type I configuration (see the inset of Fig. 4), $\Lambda$=0. While we have $\Lambda$=1 and -1 for type II and type III configurations.  The total energy of a gold and Teflon nanoplate are shown in Figs. 7(a) and 7(b), respectively. The minimum of $U(d)/k_BT$ corresponds to the equilibrium distance $d_c$. Clearly, stable quantum trapping can be realized for a gold (Teflon) nanoplate when VO$_2$ is in the metallic (insulating) phase. Due to the balance of repulsive Casimir force and gravity, stable trapping can also be realized for type II configuration.  Theoretically, the transition rate from the equilibrium distance to another position due to the Brownian motion is proportional to $\exp(-\triangle U/k_BT)$ \cite{Pha:12, Rod:10b}, where $\triangle U$ represents the energy barrier between these two positions.  The calculated results indicate that the transition rates from Casimir equilibria to stiction are negligible since the energy barriers $\triangle U//k_BT$ are quite large (e.g., over $10^4$) for the gold and Teflon nanoplates. For a flipped-down system (type III), quantum trapping can be realized for gold (Teflon) nanoplate when VO$_2$ is in the metallic (insulating) phase. However, there is a nonzero possibility that the nanoplates can escape from the equilibrium distances to the free-liquid regime ($d\rightarrow\infty$). Fortunately, the energy barrier $\triangle U/k_BT$ for such a transition is the order of $10^2$ as shown in Figs. 7(a) and 7(b), and the transition rate of the escape is also negligible.

\section{Conclusions}

In summary, the Casimir forces between a nanoplate and a layered structure
containing VO$_2$ films are investigated. In a liquid-separated environment, not only the magnitude of Casimir forces can be modified, but also the sign could be switched (e.g., from attraction to repulsion), due to the phase-transition of VO$_2$. Moreover, a stable Casimir suspension of nanoplates and its tunability are revealed. For a gold nanoplate, a switch from the quantum
trapping to its release is obtained under the metal-to-insulator transition
of VO$_2$. In addition, the quantum trapping and release of a Teflon
nanoplate are demonstrated with a complementary design. The switching
performances due to the layer thicknesses, gravitation and temperatures are
discussed as well. Theoretically, the bromobenzene can be substituted by
other high-refractive-index liquids (e.g., glycerol and styrene \cite{Van:10}%
) as long as the boiling points are larger than $T_c$. The Teflon can also
be replaced by other low-refractive-index materials (e.g., mesoporous silica
\cite{Dou:14}). This work offers the possibility of designing switchable
devices in MEMS/NEMS, resulting from the quantum fluctuations of the
electromagnetic field.

\begin{acknowledgments}
This work is supported by the National Natural Science Foundation of China
(Grant No. 11804288, No. 11704254, No. 61571386 and No. 61974127), and the
Innovation Scientists and Technicians Troop Construction Projects of Henan
Province. The research of L.X. Ge is further supported by Nanhu Scholars
Program for Young Scholars of XYNU.
\end{acknowledgments}

\appendix

\section{The permittivity of gold}

Here, a generalized Drude-Lorentz
model is applied for the permittivity of gold \cite{Seh:17}:
\begin{equation}
\varepsilon (i\xi )=\varepsilon _{D}(i\xi )+\varepsilon _{L}(i\xi ),
\end{equation}%
where the Drude term is given by:%
\begin{equation}
\varepsilon _{D}(i\xi )=\varepsilon _{\infty }+\frac{\gamma \sigma }{\xi
(\xi +\gamma )},
\end{equation}
where $\varepsilon _{\infty }=$0.83409, $\sigma =$3134.5 eV, and  $\gamma =$0.02334 eV. The Lorentz term is described by four pairs of poles:
\begin{equation}
\varepsilon_L(i\xi )=\overset{4}{\underset{j=1}{\sum }}\left( \frac{i\sigma
_{j}}{i\xi -\Omega _{j}}+\frac{i\sigma _{j}^{\ast }}{i\xi +\Omega _{j}^{\ast
}}\right)
\end{equation}%
where $\sigma _{j}$ and $\Omega _{j}$ are the generalized conductivity and resonant frequency of the $j$-th Lorentz pole. The star superscripts represent the operation of complex conjugation. The generalized Drude-Lorentz model respects causality, and it can represent the exact physical resonances in the material. The parameters for the model are listed in the Table I.

\begin{table}[ht]
\caption{The fitted parameters for Lorentz poles of gold \protect\cite{Seh:17}.}
\label{table:gold}% title of Table
\centering % used for centering table
\begin{tabular}{p{1cm}p{3cm}p{3cm}}
\hline\hline
$j$-th & $\sigma_j (\mathrm{eV})$ & $\Omega_j (\mathrm{eV})$ \\[1.5ex] \hline
1 & -0.01743+0.3059*I & 2.6905-0.16645*I \\[1.5ex]
2 & 1.0349+1.2919*I & 2.8772-0.44473*I \\[1.5ex]
3 & 1.2274+2.5605*I & 3.7911-0.81981*I \\[1.5ex]
4 & 9.85+37.614*I & 4.8532-13.891*I \\[1.0ex] \hline
\end{tabular}
% is used to refer this table in the text
\end{table}

\section{The permittivity of VO$_{2}$}

For temperature $T>T_{c}$, VO$_{2} $ is in the metallic phase, and the permittivity is given by \cite{Pir:08,Cas:07}
\begin{eqnarray}
\varepsilon (i\xi ) &=&1+\frac{\omega _{p}^{2}}{\xi (\xi +\gamma )}+\frac{%
\varepsilon _{\infty }-1}{1+\xi ^{2}/\omega _{\infty }^{2}}  \notag \\
&&+\underset{j=1}{\overset{4}{\sum }}\frac{s_{j}}{1+(\xi /\omega
_{j})^{2}+\Gamma _{j}\xi /\omega _{j}},
\end{eqnarray}%
where $\varepsilon _{\infty }=3.95,\omega _{p}=3.33$ eV, and $\gamma =0.66$
eV. The parameters $s_{j}$
and $\Gamma _{j}$ represent respectively the strength and linewidth of the $%
j $-th oscillator (resonant frequency $\omega _{j}$). 

For temperature $T<T_{c}$, VO$_{2} $ is in the insulating phase, and the permittivity is described as
\begin{equation}
\varepsilon (i\xi )=1+\frac{\varepsilon _{\infty }-1}{1+\xi ^{2}/\omega
_{\infty }^{2}}+\underset{j=1}{\overset{7}{\sum }}\frac{s_{j}}{1+(\xi
/\omega _{j})^{2}+\Gamma _{j}\xi /\omega _{j}},
\end{equation}%
where $\varepsilon _{\infty }=4.26$ and $\omega _{\infty }=15$ eV. The above equations for metallic and insulating VO$_2$ are valid for a wide range of frequency (up to about 10 eV)\cite{Cas:07}, which are modified versions of Ref. \cite{Ver:68}. The parameters are listed in Table II.

\begin{table}[ht]
\caption{The parameters for the metallic and insulating VO$_2$ \cite{Cas:07}.}
\label{table:insulating}% title of Table
\centering % used for centering table
\begin{tabular}{p{2cm}p{1.5cm}p{1.5cm}p{1.5cm}}
\hline\hline
$j$-th ($T>T_c$) & $S_j$ & $\omega_j(\mathrm{eV})$ & $\Gamma_j$ \\[0.5ex] \hline
1 & 1.816 & 0.86 & 0.95 \\[1ex]
2 & 0.972 & 2.8 & 0.23 \\[1ex]
3 & 1.04 & 3.48 & 0.28 \\[1ex]
4 & 1.05 & 4.6 & 0.34 \\[3ex]\hline\hline
$j$-th ($T<T_c$) & $S_j$ & $\omega_j(\mathrm{eV})$ & $\Gamma_j$ \\[0.5ex] \hline
1 & 0.79 & 1.02 & 0.55 \\[1ex]
2 & 0.474 & 1.30 & 0.55 \\[1ex]
3 & 0.483 & 1.50 & 0.50 \\[1ex]
4 & 0.536 & 2.75 & 0.22 \\[1ex]
5 & 1.316 & 3.49 & 0.47 \\[1ex]
6 & 1.060 & 3.76 & 0.38 \\[1ex]
7 & 0.99 & 5.1 & 0.385 \\[1ex] \hline
\end{tabular}
% is used to refer this table in the text
\end{table}

\begin{table}[ht]
\caption{The parameters for Teflon(left) and bromobenzene (right)\cite{Van:10}.}
\label{table:nonlin}% title of Table
\centering % used for centering table
\begin{tabular}{p{1.2cm}p{1.5cm}p{1.5cm}|p{1.5cm}p{1.5cm}}
\hline\hline
$j$-th & $C_j$ & $\omega_j(\mathrm{eV})$ & $C_j$ & $\omega_j (\mathrm{eV})$\\[0.5ex] \hline
1 & 0.0093 & 0.0003 & 0.0544 & 0.00502\\[1ex]
2 & 0.0183 & 0.0076 & 0.0184 & 0.0309\\[1ex]
3 & 0.139 & 0.0557 & 0.0475 & 0.111\\[1ex]
4 & 0.112 & 0.126 & 0.532 & 6.75\\[1ex]
5 & 0.195 & 6.71 & 0.645 & 13.3\\[1ex]
6 & 0.438 & 18.6 & 0.240 & 24.0 \\[1ex]
7 & 0.106 & 42.1 & 0.00927 & 99.9\\[1ex]
8 & 0.0386 & 77.6 \\[1ex] \hline
\end{tabular}
% is used to refer this table in the text
\end{table}

\section{The permittivity of Teflon and bromobenzene}

The permittivity for the Teflon and bromobenzene are given by the oscillator
model \cite{Van:10}:
\begin{equation}
\varepsilon (i\xi )=1+\underset{j}{\overset{}{\sum }}\frac{C_{j}}{1+(\xi
/\omega _{j})^{2}},
\end{equation}%
where $C_{j}$ corresponds to the oscillator strength for the $j$-th
resonance, and $\omega _{j}$ is the corresponding resonant frequency. The
values of $C_{j}$ and $\omega _{j}$ listed in Table III are fitted from the experimental data in a wide range of frequency.

\bigskip

%\begin{thebibliography}{99}
%\end{thebibliography}
%\bibliographystyle{plain}
\bibliography{references}

\end{document}